\documentclass[12pt]{article}%
\usepackage[a4paper, margin=1.25in]{geometry}
\usepackage{latexsym,epsfig}
\usepackage{epsfig,graphicx}
\usepackage{booktabs}
\usepackage{amsfonts}
\usepackage{tabularx}
\usepackage{rotating}
\usepackage{hyperref}
\usepackage{mathrsfs}
\usepackage{amsmath}
\usepackage{amssymb}
\usepackage{graphicx}
\usepackage{color}
\usepackage{natbib}
\usepackage[normalem]{ulem}
\usepackage{appendix}
\usepackage{natbib}
\usepackage[pscoord]{eso-pic}
\usepackage{fancyvrb}
\usepackage{graphicx}
\usepackage{caption}
\usepackage{subcaption}
\usepackage{float}
\usepackage{xr}
\usepackage{pgffor}
\usepackage{xstring}%
\usepackage[nodisplayskipstretch]{setspace}
\providecommand{\U}[1]{\protect\rule{.1in}{.1in}}
\newtheorem{theorem}{Theorem}

\newtheorem{corollary}{Corollary}

\newtheorem{lemma}{Lemma}

\newtheorem{proposition}{Proposition}

\graphicspath{{figures/}{figures/without_perturbation}
{figures/without_perturbation/group_1_figures_main_text/}
{figures/without_perturbation/group_1_figures_supplement/}
{figures/without_perturbation/group_2_figures_main_text/}
{figures/without_perturbation/group_2_figures_supplement/}
{figures/without_perturbation/group_3_figures_main_text/}
{figures/without_perturbation/group_3_figures_supplement/}
{figures/with_perturbation}
{figures/with_perturbation/group_1_figures_main_text/}
{figures/with_perturbation/group_1_figures_supplement/}
{figures/with_perturbation/group_2_figures_main_text/}
{figures/with_perturbation/group_2_figures_supplement/}
{figures/with_perturbation/group_3_figures_main_text/}
{figures/with_perturbation/group_3_figures_supplement/}}

\begin{document}

\title{Power Bounds and Efficiency Loss for Asymptotically Optimal Tests in IV
Regression\thanks{{\scriptsize This paper supersedes the corresponding sections
of the previous working paper ``Efficiency Loss of Asymptotically Efficient
Tests in an Instrumental Variables Regression" and the note ``Power of the CLR
Test and Total Variation Distance in an Instrumental Variables Regression" by
the same set of authors.}}}
\author{Marcelo J. Moreira 
\and Geert Ridder 
\and Mahrad Sharifvaghefi}

\maketitle

\begin{abstract}
We characterize the maximal attainable power–size gap in overidentified instrumental variables models with heteroskedastic or autocorrelated (HAC) errors. Using total variation distance and Kraft’s theorem, we define the decision theoretic frontier of the testing problem. We show that Lagrange multiplier and conditional quasi likelihood ratio tests can have power arbitrarily close to size even when the null and alternative are well separated, because they do not fully exploit the reduced-form likelihood. In contrast, the conditional likelihood ratio (CLR) test uses the full reduced-form likelihood. We prove that the power--size gap of CLR converges to one if and only if the testing problem becomes trivial in total variation distance, so that CLR attains the decision theoretic frontier whenever any test can. An empirical illustration based on Yogo (2004) shows that these failures arise in empirically relevant configurations.

\bigskip


\vspace{1.5cm}
\end{abstract}

\section{Introduction}

This paper studies the power--size tradeoff in instrumental variable models under weak identification and heteroskedastic or autocorrelated (HAC) errors. We characterize the intrinsic difficulty of the testing problem using total variation distance and derive the maximal attainable power--size gap, which defines the decision theoretic frontier for the problem. We then study whether commonly used weak-identification-robust procedures attain this frontier. In particular, we show that the conditional likelihood ratio (CLR) test attains this frontier whenever any test can do so, whereas Lagrange multiplier (LM) and conditional quasi likelihood ratio (CQLR) tests may fail to attain it in HAC environments.

Inference in instrumental variable models is complicated by weak identification. The test of \citet{AndersonRubin49} was the first procedure shown to remain valid when instruments are arbitrarily weak because its null distribution does not depend on instrument strength. However, the Anderson Rubin (AR) test is not efficient under the usual strong instrument asymptotics.

This inefficiency motivated the development of procedures that retain robustness to weak identification while achieving asymptotic efficiency when instruments are strong. Standard Wald and likelihood ratio tests are efficient under strong instruments but can exhibit severe size distortions when instruments are weak; see \citet{NelsonStartz90}, \citet{Dufour97}, \citet{StaigerStock97}, and \citet{WangZivot98}. These distortions arise because their null distributions depend on first-stage coefficients when instruments are weak. A conditioning argument resolves this problem by replacing the usual chi--square critical value with a null quantile conditional on a statistic sufficient for the first-stage coefficients, as proposed by \citet{Moreira03}. The resulting conditional Wald (CW) and conditional likelihood ratio (CLR) tests, together with the LM test, are asymptotically efficient under strong identification and valid under weak identification.

In homoskedastic models these procedures have well-understood properties. \citet{AndrewsMoreiraStock07} show that CW can exhibit severe power distortions under weak identification. In contrast, LM and CLR are unbiased and efficient under strong identification. In this setting, the efficiency gain relative to AR does not compromise asymptotic power properties, as \citet{AndrewsMoreiraStock04, AndrewsMoreiraStock06} establish that both LM and CLR are consistent. However, these results do not determine whether existing procedures fully exploit the information available in more general environments.

The situation changes in overidentified models with heteroskedastic or autocorrelated errors. In these models the reduced-form covariance structure contains information that is not fully exploited by tests constructed from AR and LM statistics. Tests designed to improve upon AR can therefore lose power for reasons that do not arise under homoskedasticity. We show that this loss of information can be structural rather than local.

In instrumental variable models with heteroskedastic or autocorrelated errors the reduced-form covariance matrix introduces many nuisance parameters. When the number of instruments increases, the number of covariance parameters grows roughly with the square of the number of instruments, making the testing problem inherently high dimensional. Classical power comparisons, such as those of \citet{AndrewsMoreiraStock06}, depend on a small number of parameters even when the number of instruments increases. In contrast, the HAC IV framework allows a much richer set of covariance configurations. Showing that power failures arise in such a rich environment is therefore not straightforward.

Our first contribution is to characterize the maximal attainable power--size gap using the total variation distance between the null and alternative distributions. \citet{BertanhaMoreira20} show that total variation distance is the relevant metric for determining whether any test can deliver nontrivial power. By \citet{Kraft55}, the minimum total variation distance between the convex hulls of the null and alternative models determines the largest power--size gap achievable by any measurable test. We derive an explicit bound for this frontier in the linear IV model and show that it provides a necessary and sufficient condition for asymptotic distinguishability.

We then show that the power--size gap of the conditional likelihood ratio (CLR) test converges to one if and only if the decision theoretic frontier converges to one. Thus, CLR attains the maximal attainable power--size gap whenever any test can do so. This result reflects the fact that CLR exploits the full reduced-form information available in the HAC model.

It is useful to contrast this property with existing procedures. AR also attains the frontier, although it is not efficient under strong identification. The LM test and the CQLR test were developed to improve upon AR while preserving robustness to weak identification. The CQLR test can be viewed as an HAC extension of CLR following \citet{Kleibergen05}. However, the two procedures are not equivalent in HAC environments. CQLR is constructed from statistics based on AR and LM components, whereas CLR is defined directly from the Gaussian reduced-form likelihood; see \citet{AndrewsMikusheva16} and \citet{MoreiraMoreira19}. Consequently, CQLR may discard information contained in the reduced-form covariance structure that CLR fully exploits.

We identify a class of data generating processes, which we call \emph{impossibility designs}, under which the LM noncentrality parameter remains bounded even when the null and alternative distributions are arbitrarily well separated. In such designs LM and CQLR can be inconsistent, with power arbitrarily close to size. This phenomenon has no analog in the homoskedastic case.

These designs are not pathological. The set of data generating processes that generate bounded noncentrality parameters has positive measure, and the associated power losses extend to neighborhoods of these designs. Using data of \citet{Yogo04}, we show that empirically plausible parameter configurations can lie near this region. In such cases AR--LM based conditional procedures can suffer substantial power losses, whereas CLR maintains high power.

Taken together, our results establish a sharp separation between CLR and both LM and CQLR tests in instrumental variable models with HAC errors. CLR exploits the full reduced-form likelihood and achieves the decision theoretic frontier, whereas LM and CQLR operate in a restricted information space and may fail to convert statistical separation into power.

\citet{Dufour97}, building on \citet{GleserHwang87}, shows that conventional Wald tests can have size arbitrarily close to one when instruments are arbitrarily weak. This finding supported the use of the AR test and the development of other procedures robust to weak instruments. Our results reveal a parallel issue for power in HAC IV models. When the covariance structure is near an impossibility design, LM and CQLR tests can have power arbitrarily close to size even though the null and alternative distributions are well separated. Just as \citet{BoundJaegerBaker95}, \citet{Dufour97}, and \citet{StaigerStock97} shifted empirical practice away from Wald tests toward procedures robust to weak instruments, our results suggest that procedures based on AR and LM statistics may discard relevant information in HAC environments. This motivates the use of procedures such as the CLR test that exploit the full reduced-form likelihood and also encourages the development of new tests that fully use the information available in the reduced-form.

The remainder of the paper is organized as follows. Section \ref{model} introduces the model and test statistics. Section \ref{sec:CLR test} studies the CLR test using total variation distance. Section \ref{sec:LMbounds} characterizes the noncentrality parameter of the LM statistic and defines impossibility designs for both LM and CQLR tests. Section \ref{regions} provides diagnostics and an empirical application. Section \ref{sec:conclusion} concludes the paper. An online supplement contains the proofs of all theoretical results, additional lemmas and a proposition, further power comparisons, and supplementary results for the empirical application.

\section{Model and test statistics}

\label{model}

We consider the instrumental variable regression model
\begin{align}
y_{1}  &  = y_{2}\beta^{\ast} + u,\\
y_{2}  &  = Z\pi+ v_{2},
\end{align}
where $y_{1}$ and $y_{2}$ are $n \times1$ vectors, $Z$ is an $n \times k$
matrix of nonrandom instruments with full column rank, and $\pi$ is a $k
\times1$ vector of first stage coefficients. For simplicity we omit additional covariates. They can be partialled out using orthogonal projections according to the Frisch–Waugh–Lovell (FWL) theorem; see \citet{AndrewsMoreiraStock06} for the IV model. The disturbances $u$ and $v_{2}$
have mean zero and may exhibit heteroskedasticity and autocorrelation.

Our objective is to test
\[
H_{0}: \beta^{\ast} = \beta_{0} \quad\text{against}\quad H_{1}: \beta^{\ast}
\neq\beta_{0},
\]
treating $\pi$ as a nuisance parameter.

Let $Y=[y_{1}\ y_{2}]$. The reduced form can be written as
\begin{equation}
Y = Z\pi a^{\ast\prime} + V,
\end{equation}
where $a^{\ast}=(\beta^{\ast},1)^{\prime}$ and $V=[v_{1}\ v_{2}]$ with
$v_{1}=u+\beta^{\ast}v_{2}$.

Define
\begin{equation}
R = (Z^{\prime}Z)^{-1/2}Z^{\prime}Y = \mu a^{\ast\prime} + \widetilde{V},
\end{equation}
where $\mu=(Z^{\prime}Z)^{1/2}\pi$ and $\widetilde{V}=(Z^{\prime}%
Z)^{-1/2}Z^{\prime}V$.

We impose the following normalized average conditions.

\bigskip

\noindent\textbf{Assumption NA.}

\begin{enumerate}
\item[(a)] $n^{-1}Z^{\prime}Z \to D$, where $D$ is positive definite.

\item[(b)] $n^{-1}V^{\prime}V \overset{p}{\to} \Omega$, where $\Omega$ is
positive definite.

\item[(c)] $(Z^{\prime}Z)^{-1/2}Z^{\prime}V \overset{d}{\to} N(0,\Sigma)$,
where $\Sigma$ is positive definite.
\end{enumerate}

Under Assumption NA, the finite dimensional distribution of $R$ converges to a
Gaussian limit experiment with covariance matrix $\Sigma$. This limit
experiment coincides with the weak instrument asymptotics of
\citet{StaigerStock97} and provides a sharp approximation to finite-sample
behavior when first-stage coefficients are local to zero. It also forms the
basis for our fixed-alternative comparisons in later sections.

Accordingly, we conduct our analysis under the Gaussian reduced-form model
\begin{equation}
R \sim N\!\left(  \mu a^{\ast\prime}, \Sigma\right) ,
\end{equation}
treating $\Sigma$ as known. This formulation isolates the information content
of the reduced form and allows us to analyze the geometry of the testing
problem directly.

Fix $\beta_{0}$ and define
\[
a_{0}=(\beta_{0},1)^{\prime}, \qquad b_{0}=(1,-\beta_{0})^{\prime}.
\]
Define
\begin{align}
T  &  = \left[  (a_{0}^{\prime}\otimes I_{k})\Sigma^{-1} (a_{0}\otimes
I_{k})\right] ^{-1/2} (a_{0}^{\prime}\otimes I_{k})\Sigma^{-1}\mathrm{vec}%
(R),\\
S  &  = \left[  (b_{0}^{\prime}\otimes I_{k})\Sigma(b_{0}\otimes I_{k})\right]
^{-1/2} (b_{0}^{\prime}\otimes I_{k})\mathrm{vec}(R).
\end{align}
Under $H_{0}$, $T$ is sufficient for $\mu$, while $S$ is pivotal and
independent of $T$.

Let
\[
B_{0}=
\begin{pmatrix}
1 & 0\\
-\beta_{0} & 1
\end{pmatrix}
, \qquad R_{0}=RB_{0}.
\]
Then
\[
\mathrm{vec}(R_{0}) \sim N\!\left(  \mathrm{vec}(\mu a_{\Delta}^{\prime}),
\Sigma_{0} \right) ,
\]
where $a_{\Delta}=(\Delta,1)^{\prime}$ with $\Delta=\beta^{\ast}-\beta_{0}$
and
\[
\Sigma_{0} = (B_{0}^{\prime}\otimes I_{k}) \Sigma(B_{0}\otimes I_{k}) =
\begin{pmatrix}
\Sigma_{11} & \Sigma_{12}\\
\Sigma_{21} & \Sigma_{22}%
\end{pmatrix}
.
\]

The Anderson Rubin statistic is
\begin{equation}
AR=S^{\prime}S.\label{AR}%
\end{equation}

The one-sided and two-sided LM statistics are
\begin{equation}
LM_{1} = \frac{S^{\prime}\Sigma_{11}^{-1/2}(\Sigma^{22})^{-1/2}T} {\left\Vert
\Sigma_{11}^{-1/2}(\Sigma^{22})^{-1/2}T \right\Vert }, \qquad LM = \frac{
\left(  S^{\prime}\Sigma_{11}^{-1/2}(\Sigma^{22})^{-1/2}T \right) ^{2} } {
T^{\prime} (\Sigma^{22})^{-1/2}\Sigma_{11}^{-1} (\Sigma^{22})^{-1/2}T
},\label{eqLM1}%
\end{equation}
where
\begin{equation}\label{eq:sigma22}
\Sigma^{22} = (\Sigma_{22} - \Sigma_{21}\Sigma_{11}^{-1}\Sigma_{12})^{-1}.
\end{equation}

Another commonly used statistic is the quasi likelihood ratio (QLR):
\begin{equation}
QLR = \frac{ AR-r(T)+ \sqrt{(AR-r(T))^{2}+4\,LM\,r(T)} }{2},\label{QLR}%
\end{equation}
where natural choices for the rank statistic $r(T)$ are
\begin{align*}
r_{1}(T)  &  =T^{\prime}T \text{ and } \\
r_{2}(T)  &  =T^{\prime} (\Sigma^{22})^{-1/2} \Sigma_{11}^{-1} (\Sigma
^{22})^{-1/2} T,
\end{align*}
as outlined by \citet{AndrewsGuggenberger17}. We refer to these variants as $QLR1$ and $QLR2$.

The null distribution of $QLR$ depends on $\mu$. Since $T$ is sufficient for
$\mu$ under $H_{0}$, we follow the conditioning argument of \citet{Moreira03}
and define conditional versions that reject the null hypothesis when the statistic exceeds its
null quantile conditional on $T$. The conditional critical value function
based on a statistic $\psi$ is
\begin{equation}
c_{\alpha}(T) = \min\left\{  c: \sup_{\Delta=0,\;\mu\neq0} \Pr\!\left(  \psi>c
\mid T \right)  \le\alpha\right\} .\label{eq:conditional_cvf}%
\end{equation}

We refer to this as the conditional QLR (CQLR) test. \citet{Andrews16} shows
that CQLR is a special case of conditional linear combination (CLC) tests of
the form
\begin{equation}
LC = w(T)\cdot AR + (1-w(T))\cdot LM,\label{CLC}%
\end{equation}
where $0\le w(T)\le1$ and proposes the plug-in conditional linear combination test (PI-CLC).

Finally, following \citet{Moreira03} for the homoskedastic case,
\citet{AndrewsMikusheva16} and \citet{MoreiraMoreira19} define the likelihood ratio statistic for the HAC case:
\begin{equation}
LR = Q(\beta_{0}) - \inf_{\beta\in\mathbb{R}}Q(\beta),\label{eq:LR_stat}%
\end{equation}
where
\begin{equation}
Q(\beta) = \mathrm{vec}(R)^{\prime} (b\otimes I_{k}) \left[  (b^{\prime
}\otimes I_{k}) \Sigma(b\otimes I_{k}) \right] ^{-1} (b^{\prime}\otimes I_{k})
\mathrm{vec}(R),
\end{equation}
with $b=(1,-\beta)^{\prime}$.

The conditional likelihood ratio (CLR) test rejects when $LR$ exceeds its null
quantile conditional on $T$.

CQLR extends the algebraic structure of CLR from the homoskedastic setting to the GMM framework with heteroskedastic or autocorrelated errors, whereas CLR is defined directly from the Gaussian reduced-form likelihood. The two procedures coincide under homoskedasticity but are generally distinct in HAC environments.

When $k>2$ in HAC settings, the likelihood ratio statistic does not admit a
closed-form solution to the minimization problem in \eqref{eq:LR_stat}. \citet{MoreiraNeweySharifvaghefi24} establish this result and develop computational algebra methods that allow the CLR statistic to be computed reliably in such environments.

\section{CLR test}

\label{sec:CLR test}

In this section, we show that the power--size gap of the CLR test, which
rejects the null hypothesis when the $LR$ statistic defined in
\eqref{eq:LR_stat} exceeds its conditional critical value given by
\eqref{eq:conditional_cvf}, converges to one as the minimum total variation
distance between the convex hulls of the distributions under the null and
alternative hypotheses for $R_{0}$ converges to one. By \citet{Kraft55}'s
theorem, discussed below, this result implies that the power--size gap of any
test can converge to one only if the same holds for the CLR test.

To place \citet{Kraft55}'s theorem in the context of linear IV models, let
$\phi_{\Delta,\mu}(\cdot)$ denote the normal probability density function of
$\mathrm{vec}(R_{0})$ with mean $\mathrm{vec}(\mu a_{\Delta}^{\prime})$ and
variance matrix $\Sigma_{0}$. Denote the convex hull of the set of normal
densities under the null hypothesis by
\begin{equation}
\label{eq:null_hypothesis}\mathcal{C}_{0} = \left\{  \sum_{i=1}^{m}
c_{i}\,\phi_{0,\mu_{i}}(\cdot) : c_{i}\ge0,\; \sum_{i=1}^{m} c_{i}=1,\;
m\in\mathbb{N} \right\} .
\end{equation}

Similarly, let
\begin{equation}
\label{eq:alt_hypothesis}\mathcal{C}_{1} = \left\{  \sum_{j=1}^{m} c_{j}%
\,\phi_{\Delta_{j},\mu_{j}}(\cdot) : c_{j}\ge0,\; \sum_{j=1}^{m} c_{j}=1,\;
\Delta_{j}^{2}\mu_{j}^{\prime}\Sigma_{11}^{-1}\mu_{j}\ge d,\; m\in\mathbb{N}
\right\}
\end{equation}
denote the convex hull of the set of normal densities under the alternative
hypothesis satisfying
\begin{equation}
\Delta^{2}\mu^{\prime}\Sigma_{11}^{-1}\mu\ge d,
\end{equation}
where $d>0$ is an arbitrary constant.

The separation condition in \eqref{eq:alt_hypothesis} reflects the intrinsic
statistical distance between the null and alternative distributions in the
Gaussian reduced-form model. The scalar
\[
\Delta^{2}\mu^{\prime}\Sigma_{11}^{-1}\mu
\]
is central for three reasons. First, it equals the noncentrality parameter of
the Anderson--Rubin statistic under the alternative, measuring the structural
deviation $\Delta$ scaled by instrument strength as summarized by
$\mu^{\prime}\Sigma_{11}^{-1}\mu$. Thus alternatives satisfying
$\Delta^{2}\mu^{\prime}\Sigma_{11}^{-1}\mu \ge d$ are separated from the null
by at least a fixed AR signal. Second, as shown below, a useful bound for the total
variation distance over nuisance mean vectors under the null depends only on
this scalar, so it directly indexes intrinsic distinguishability between the
null and alternative models. Third, it has a natural econometric
interpretation: $\Delta^{2} = (\beta^{\ast}-\beta_{0})^{2}$ is the squared
structural distance from the null, while $\mu^{\prime}\Sigma_{11}^{-1}\mu$
summarizes instrument strength. Their product therefore measures structural
separation scaled by instrument relevance. For these reasons, defining the
alternative convex hull through
$\Delta^{2}\mu^{\prime}\Sigma_{11}^{-1}\mu \ge d$ provides a model-based and
decision theoretic notion of separation that allows a sharp characterization
of the maximal attainable power--size gap.

The total variation (TV) distance between $f(\cdot)\in\mathcal{C}_{0}$ and
$g(\cdot)\in\mathcal{C}_{1}$ is defined as
\begin{equation}
\label{eq:TV_distance}D_{\mathrm{TV}}(f,g) = \frac{1}{2}\int_{\mathbb{R}^{2k}}
\left| f(r_{0})-g(r_{0}) \right|\,dr_{0}.
\end{equation}

\citet{Kraft55} shows that, in order for there to exist a test
$\psi(\cdot):\mathbb{R}^{2k}\to[0,1]$ such that
\begin{equation}
\inf_{g\in\mathcal{C}_{1}}\mathbb{E}_{g} \left(\psi(R_{0})\right) - \sup
_{f\in\mathcal{C}_{0}}\mathbb{E}_{f} \left(\psi(R_{0})\right) \ge q,
\end{equation}
it is necessary and sufficient that
\begin{equation}
\min_{\substack{f\in\mathcal{C}_{0}\\g\in\mathcal{C}_{1}}} D_{\mathrm{TV}}
(f,g)=q.
\end{equation}
That is, the power--size gap of a test can be at least $q$ if and only if the
minimum total variation distance between $\mathcal{C}_{0}$ and
$\mathcal{C}_{1}$ equals $q$\footnote{The proof of this result is nonconstructive: it relies on an infinite-dimensional separating hyperplane argument that establishes the existence of a test achieving the bound but does not provide an explicit form for such a test.}. Consequently, a power--size gap equal to $1$ is possible if and only if the minimum total variation distance equals $1$.

Total variation distance plays a central role in our analysis because it is
the decision theoretic metric that determines what any hypothesis test can
accomplish. \citet{Kraft55}'s theorem is sharp: the largest power--size gap
attainable by any measurable test, including randomized tests, coincides with
the minimal TV distance between the convex hulls of the null and alternative
models. Other notions of distance, such as the L\'evy--Prokhorov distance, can
be useful for specific arguments but, as \citet{BertanhaMoreira20} show, they
characterize only more restricted classes of tests and therefore do not
capture the full difficulty of the testing problem. For this reason, TV
distance provides the natural benchmark for the intrinsic difficulty of
distinguishing the null from the alternative. Using this metric, we derive convenient bounds showing that the power of the CLR test approaches the decision
theoretic frontier when the null and alternative distributions become well
separated in TV, a property that AR-LM-based procedures cannot generally
replicate in HAC environments.

The minimum total variation distance between $\mathcal{C}_{0}$ and $\mathcal{C}_{1}$ is smaller than the minimum total variation distance between 
the extreme points of these sets. Therefore,
\begin{equation}
\min_{\substack{f\in\mathcal{C}_{0}\\g\in\mathcal{C}_{1}}} D_{\mathrm{TV}%
}(f,g) \leq \min_{\substack{\mu_{i},\mu_{j},\Delta\\\text{s.t. } \Delta^{2}%
\mu_{j}^{\prime}\Sigma_{11}^{-1}\mu_{j}\ge d}} D_{\mathrm{TV}} \left(\phi
_{0,\mu_{i}},\phi_{\Delta,\mu_{j}}\right).
\end{equation}

It can be shown that
\begin{equation}
\label{eq:TV_normal}D_{\mathrm{TV}} \left(\phi_{0,\mu_{i}},\phi_{\Delta
,\mu_{j}}\right) = 2\Phi\!\left(  \frac{\delta(\mu_{i},\mu_{j},\Delta)}%
{2}\right)  -1 = F_{\chi^{2}_{1}}\!\left(  \frac{\delta^{2}(\mu_{i},\mu
_{j},\Delta)}{4}\right) ,
\end{equation}
where $\Phi(\cdot)$ and $F_{\chi^{2}_{1}}(\cdot)$ denote the cumulative
distribution functions of a standard normal random variable and a chi--square
random variable with one degree of freedom, respectively. Moreover,
\begin{equation}
\delta^{2}(\mu_{i},\mu_{j},\Delta) = \Delta^{2}\mu_{j}^{\prime}\Sigma^{11}%
\mu_{j} + 2\Delta(\mu_{j}-\mu_{i})^{\prime}\Sigma^{21}\mu_{j} + (\mu_{j}%
-\mu_{i})^{\prime}\Sigma^{22}(\mu_{j}-\mu_{i}),
\end{equation}
where $\Sigma^{22}$ is given by \eqref{eq:sigma22},
\[
\Sigma^{11} = \left(  \Sigma_{11}-\Sigma_{12}\Sigma_{22}^{-1}\Sigma
_{21}\right) ^{-1}, \qquad \text{and} \qquad\Sigma^{21} = -\Sigma
^{22}\Sigma_{21}\Sigma_{11}^{-1}.
\]
Therefore,
\begin{equation}
\min_{\substack{f\in\mathcal{C}_{0}\\g\in\mathcal{C}_{1}}} D_{\mathrm{TV}%
}(f,g) \leq \min_{\substack{\mu_{i},\mu_{j},\Delta\\\text{s.t. } \Delta^{2}%
\mu_{j}^{\prime}\Sigma_{11}^{-1}\mu_{j}\ge d}} F_{\chi^{2}_{1}}\left(
\frac{\delta^{2}(\mu_{i},\mu_{j},\Delta)}{4} \right) .
\end{equation}
Since $F_{\chi^{2}_{1}}(\cdot)$ is increasing, we can write
\begin{equation}
\min_{\substack{f\in\mathcal{C}_{0}\\g\in\mathcal{C}_{1}}} D_{\mathrm{TV}%
}(f,g) \leq F_{\chi^{2}_{1}}\left(  \frac{\delta_{\min}^{2}}{4}\right) ,
\end{equation}
where
\begin{equation}
\delta_{\min}^{2} = \min_{\substack{\mu_{i},\mu_{j},\Delta\\\text{s.t. }
\Delta^{2}\mu_{j}^{\prime}\Sigma_{11}^{-1}\mu_{j}\ge d}} \delta^{2}(\mu
_{i},\mu_{j},\Delta).
\end{equation}
The solution to this constrained optimization problem is $\delta_{\min}^{2}%
=d$. Consequently,
\begin{equation}
\min_{\substack{f\in\mathcal{C}_{0}\\g\in\mathcal{C}_{1}}} D_{\mathrm{TV}%
}(f,g) \leq F_{\chi^{2}_{1}}\left(  \frac{d}{4}\right) .
\end{equation}
Lemma \ref{lem:min_TV_distance} formalizes this finding.

\begin{lemma}
\label{lem:min_TV_distance} Let $\mathcal{C}_{0}$, as defined in
\eqref{eq:null_hypothesis}, be the convex hull of distributions under the null
hypothesis, and let $\mathcal{C}_{1}$, as defined in
\eqref{eq:alt_hypothesis}, be the convex hull of distributions under the
alternative hypothesis. Consider the total variation distance between
$f(\cdot)\in\mathcal{C}_{0}$ and $g(\cdot)\in\mathcal{C}_{1}$ as defined in
\eqref{eq:TV_distance}. Under Assumption NA,
\[
\min_{\substack{f\in\mathcal{C}_{0}\\g\in\mathcal{C}_{1}}} D_{\mathrm{TV}%
}(f,g) \leq F_{\chi^{2}_{1}}\left(  \frac{d}{4}\right) .
\]

\end{lemma}

$F_{\chi^{2}_{1}}\left(d/4\right) $ is always less than one and converges to one if and only if $d \to \infty$. Therefore, as the total variation distance between $\mathcal{C}_{0}$ and $\mathcal{C}_{1}$ converges to one,
$d\to\infty$. In the next step, we show that as $d\to\infty$, the power--size
gap of the CLR test converges to one. Consequently, \citet{Kraft55}'s theorem
implies that the power--size gap of any test can converge to one only if the
power--size gap of the CLR test does.

Finding the exact power of the CLR test is complicated because the $LR$
statistic depends on an optimization problem with no closed-form solution.
Moreover, the critical value function of the CLR test is random. To address
these issues, we consider a lower bound for CLR power.

Since $\inf_{\beta\in\mathbb{R}}Q(\beta)\ge0$, the conditional critical value
function defined in \eqref{eq:conditional_cvf} satisfies
\begin{equation}
c_{\alpha}(T) \le\min\left\{  c: \sup_{\Delta=0,\;\mu\neq0} \Pr\!\left(
Q(\beta_{0})>c \,\middle|\, T \right)  \le\alpha\right\} .
\end{equation}
Given that $Q(\beta_{0})=S^{\prime}S$ is independent of $T$, we can further
write
\begin{equation}
c_{\alpha}(T) \le\min\left\{  c: \sup_{\Delta=0,\;\mu\neq0} \Pr\!\left(
Q(\beta_{0})>c \right)  \le\alpha\right\} .
\end{equation}
Under the null hypothesis $\Delta=0$, $Q(\beta_{0})$ follows a chi--square
distribution with $k$ degrees of freedom. Therefore,
\begin{equation}
c_{\alpha}(T)\le q_{1-\alpha}(k),
\end{equation}
where $q_{1-\alpha}(k)$ denotes the $(1-\alpha)$ quantile of a chi--square
distribution with $k$ degrees of freedom. Lemma \ref{lem:cv_bound} states this
upper bound.

\begin{lemma}
\label{lem:cv_bound} Under Assumption NA and for a given nominal size $\alpha
$, the conditional critical value function $c_{\alpha}(T)$ of the CLR test,
defined in \eqref{eq:conditional_cvf}, is bounded above by $q_{1-\alpha}(k)$.
\end{lemma}

It follows by Lemma \ref{lem:cv_bound} that
\begin{equation}
\label{eq:power_lower_bound_cont}\Pr\!\left[  Q(\beta_{0})-\inf_{\beta
\in\mathbb{R}}Q(\beta)>c_{\alpha}(T) \right]  \ge\Pr\!\left[  Q(\beta
_{0})-\inf_{\beta\in\mathbb{R}}Q(\beta)>q_{1-\alpha}(k) \right] .
\end{equation}
Moreover, since $\inf_{\beta\in\mathbb{R}}Q(\beta)\le Q(\beta^{\ast})$,
\begin{equation}
\Pr\!\left[  Q(\beta_{0})-\inf_{\beta\in\mathbb{R}}Q(\beta)>q_{1-\alpha}(k)
\right]  \ge\Pr\!\left[  Q(\beta_{0})-Q(\beta^{\ast})>q_{1-\alpha}(k) \right]
.
\end{equation}
Thus, the CLR power is bounded from below by
\[
\Pr\!\left[  Q(\beta_{0})-Q(\beta^{\ast})>q_{1-\alpha}(k) \right] .
\]

We have
\begin{equation}
Q(\beta^{\ast}) = \mathrm{vec}(R)^{\prime}(b\otimes I_{k}) \left[  (b^{\prime
}\otimes I_{k})\Sigma(b\otimes I_{k}) \right] ^{-1} (b^{\prime}\otimes I_{k})
\mathrm{vec}(R) \sim\chi^{2}(k).
\end{equation}
Moreover, $Q(\beta_{0})=S^{\prime}S$ follows a noncentral chi--square
distribution with $k$ degrees of freedom and noncentrality parameter
$d=\Delta^{2}\mu^{\prime}\Sigma_{11}^{-1}\mu$. Therefore, as shown in
Proposition \ref{prop:power_lower_bound_clr_test} below, the lower bound on
CLR power converges to one at an exponential rate in $d-q_{1-\alpha}(k)$.

\begin{proposition}
\label{prop:power_lower_bound_clr_test} Consider the CLR test that rejects the
null hypothesis when the $LR$ statistic given by \eqref{eq:LR_stat} exceeds
its conditional critical value function given by \eqref{eq:conditional_cvf}.
Suppose Assumption NA holds. Then,
\begin{align}
\Pr\!\left[  LR>c_{\alpha}(T)\right]   & \geq1-2k\exp\!\left(  -\frac{1}%
{4k}\max\{0,d-q_{1-\alpha}(k)\}\right) \\
& \quad-2k\exp\!\left(  -\frac{1}{8k}\max\{0,d-q_{1-\alpha}(k)\}\right)
-2\exp\!\left(  -\frac{(\max\{0,d-q_{1-\alpha}(k)\})^{2}}{128\,d} \right) ,\nonumber
\end{align}
where $d=\Delta^{2}\mu^{\prime}\Sigma_{11}^{-1}\mu$.
\end{proposition}

To ensure the power--size gap of the CLR test converges to one as $d\to\infty
$, it suffices that $d-q_{1-\alpha}(k)\to\infty$ (for example, by choosing
critical values so that $q_{1-\alpha}(k)$ grows slower than $d$). Under this
choice, as $d\to\infty$, both the CLR power--size gap and the minimum total
variation distance between $\mathcal{C}_{0}$ and $\mathcal{C}_{1}$ converge to
one at an exponential rate in $d$. By \citet{Kraft55}, the power--size gap of
any test can converge to one only if the power--size gap of the CLR test does.
Theorem \ref{thm:power_lower_bound_clr_test} formalizes this conclusion.

\begin{theorem}
\label{thm:power_lower_bound_clr_test} Consider the CLR test with test
statistic $LR$ given by \eqref{eq:LR_stat} and conditional critical value
function $c_{\alpha}(T)$ chosen such that \eqref{eq:conditional_cvf} holds.
Suppose Assumption NA holds. Then the power--size gap of the CLR test
converges to one if and only if the minimum total variation distance between
the convex hulls of probability densities under the null and alternative
hypotheses converges to one. By \citet{Kraft55}, this implies that the
power--size gap of any test can converge to one only if the power--size gap of
the CLR test converges to one.
\end{theorem}

\section{LM-based tests and failure under impossibility designs}

\label{sec:LMbounds}

The LM test is asymptotically efficient under the conventional strong-instrument local asymptotic framework. However, this approximation can be
misleading when instruments are weak or when we consider fixed alternatives.
In such cases, the LM statistic can exhibit low power even in situations where
distinguishing the null from the alternative should be straightforward in a
decision theoretic sense. This section characterizes the behavior of the LM
statistic under fixed alternatives and shows how the resulting limitations
propagate to CLC and CQLR procedures.

We work with the one-sided LM statistic $LM_{1}$ in \eqref{eqLM1}. Under
Assumption NA, the joint normality of $(S,T)$ implies the representation
\begin{equation}
S=\Delta\Sigma_{11}^{-1/2}\mu+U_{S}, \qquad T=(\Sigma^{22})^{1/2}(I_{k}%
-\Delta\Sigma_{21}\Sigma_{11}^{-1})\mu+U_{T},
\end{equation}
where $U_{S}$ and $U_{T}$ are independent $N(0,I_{k})$ random vectors. Hence
\begin{equation}
\text{LM}_{1} = \frac{(\Delta\Sigma_{11}^{-1/2}\mu+U_{S})^{\prime} \Sigma
_{11}^{-1/2} \left(  (I_{k}-\Delta\Sigma_{21}\Sigma_{11}^{-1})\mu+(\Sigma
^{22})^{-1/2}U_{T} \right) } {\left\Vert \Sigma_{11}^{-1/2} \left(
(I_{k}-\Delta\Sigma_{21}\Sigma_{11}^{-1})\mu+(\Sigma^{22})^{-1/2}U_{T}
\right)  \right\Vert } .\label{eq:LM1_rep}%
\end{equation}

\subsection{Benchmark: strong instruments and local alternatives}

\noindent\textbf{Assumption SIV LA.} (a) $\Delta_{n}=h_{\Delta}/n^{1/2}$ for
some constant $h_{\Delta}$. (b) $\pi$ is a fixed nonzero $k$ vector.

\begin{proposition}
\label{Prop SIV-LA} Under Assumptions SIV LA and NA,
\[
LM_{1}\rightarrow_{d} N\!\left(  h_{\Delta} \left(  \pi^{\prime}D^{1/2}%
\Sigma_{11}^{-1}D^{1/2}\pi\right) ^{1/2}, 1 \right) .
\]

\end{proposition}

This is the standard efficiency result. Under strong identification and local
alternatives, the LM noncentrality increases with both the local distance from
the null and instrument strength. In this regime, the LM test is
asymptotically optimal.

\subsection{Fixed alternatives}

We now keep $\Delta$ fixed. In this regime, terms that vanish under local
asymptotics remain first order. The resulting approximation reveals that the
LM drift need not increase with $|\Delta|$.

\bigskip

\noindent\textbf{Assumption SIV FA.} (a) $\Delta$ is fixed. (b) $\pi$ is a
fixed nonzero $k$ vector.

\begin{theorem}
\label{Thm SIV-FA} Define
\begin{equation}
c(\Delta ,\mu )=\frac{\Delta \mu ^{\prime }\Sigma _{11}^{-1}\mu -\Delta
^{2}\mu ^{\prime }\Sigma _{11}^{-1}\Sigma _{21}\Sigma _{11}^{-1}\mu }{\left(
\mu ^{\prime }(I_{k}-\Delta \Sigma _{11}^{-1}\Sigma _{12})\Sigma
_{11}^{-1}(I_{k}-\Delta \Sigma _{21}\Sigma _{11}^{-1})\mu \right) ^{1/2}}.
\end{equation}
Under Assumptions SIV FA and NA,
\begin{equation*}
LM_{1}-c(\Delta ,\mu ) \rightarrow_{d} N \left( 0,\, 1+\|\gamma\|^{-2}\Delta^{2} \pi'
D^{1/2}\Sigma_{11}^{-1/2} M_{\gamma}\Sigma_{11}^{-1/2}
(\Sigma^{22})^{-1} \Sigma_{11}^{-1/2} M_{\gamma}\Sigma_{11}^{-1/2}
D^{1/2}\pi \right),
\end{equation*}
where
\begin{equation*}
\gamma =\Sigma _{11}^{-1/2}(I_{k}-\Delta \Sigma _{21}\Sigma
_{11}^{-1})D^{1/2}\pi,  \qquad \text{and} \qquad 
M_{\gamma } = I_{k} - \frac{\gamma \gamma ^{\prime }}{\Vert \gamma \Vert^{2}}.
\end{equation*}
\end{theorem}

The behavior of the statistic is driven by two forces: a stochastic component with controlled variance and a deterministic drifting term $c(\Delta, \mu )$. 

The variance does not explode when $\Delta \to \infty$. Although $\Delta$ is fixed in the present asymptotic regime, this observation highlights that the stochastic component of the statistic remains well behaved even for large values of $\Delta$.

Furthermore, the variance does not depend on $\pi$ through its norm, but only through its direction. In principle, one could characterize the smallest and largest values of this variance as $\pi$ varies, since the expression is a ratio of quadratic forms in $\pi$, but we choose not to pursue this here. The main message is that this variance is well controlled, while an important component of the statistic is the drifting term $c(\Delta,\mu)$, which need not diverge as $\Delta$ grows.

We can further decompose this asymptotic distribution into two components that correspond to the two sources of randomness in the LM statistic. The first component,
\[
N \!\left(
0,\,
1+\|\gamma\|^{-2}\Delta^{2}\pi'
D^{1/2}\Sigma_{11}^{-1}(\Sigma^{22})^{-1}\Sigma_{11}^{-1}
D^{1/2}\pi
\right),
\]
originates from the random variation in the numerator of the $LM_{1}$ statistic. The second component,
\[
N \left(
0,\,
\|\gamma\|^{-2}\Delta^{2}\pi'
D^{1/2}\Sigma_{11}^{-1/2}
\Big(
M_{\gamma}\Sigma_{11}^{-1/2}(\Sigma^{22})^{-1}\Sigma_{11}^{-1/2}M_{\gamma}
-
\Sigma_{11}^{-1/2}(\Sigma^{22})^{-1}\Sigma_{11}^{-1/2}
\Big)
\Sigma_{11}^{-1/2}D^{1/2}\pi
\right),
\]
arises from the randomness in the denominator of the $LM_{1}$ statistic.

Under both the SIV FA and SIV LA approximations, we obtain a normal approximation to $\mathrm{LM}_{1}$, rather than the more complicated mixed normal distribution that arises under weak IV approximations. Unlike the SIV LA approximation, the SIV FA approximation retains all terms required for an accurate analysis of the deterministic drifting component $c(\Delta,\mu)$ of $\mathrm{LM}_{1}$.

The drifting term $c(\Delta, \mu )$ depends on
\[
\zeta= \Delta \mu^{\prime}\Sigma_{11}^{-1} \mu- \Delta^{2}%
\mu^{\prime}\Sigma_{11}^{-1} \Sigma_{21}\Sigma_{11}^{-1}\mu,
\]
which need not have the same sign as $\Delta$. Thus, even for large $|\Delta|$, the one-sided LM test need not become powerful. This phenomenon can arise even under homoskedasticity; see \citet{AndrewsMoreiraStock06}. Although the two-sided LM statistic remains consistent under homoskedastic errors, we show below that a deeper issue arises in HAC models: the LM noncentrality parameter need not diverge even when the null and alternative distributions become well separated. Consequently, the LM statistic may fail to translate statistical separation into power.

The decomposition above highlights two distinct forces that determine the behavior of the LM statistic. The stochastic component has a well-behaved variance that remains bounded even when $\Delta$ becomes large. In contrast, the behavior of the statistic may be dominated by the deterministic drifting term $c(\Delta,\mu)$. Whether this drift becomes large depends on the structure of the quadratic form that defines $c(\Delta,\mu)$ and cannot be determined from the approximation alone. In particular, the drift need not increase with $|\Delta|$. We next study the behavior of this drifting component in detail. We show that, under HAC covariance structures, the drifting term may fail to diverge even when the null and alternative distributions become well separated.

\subsection{Impossibility designs}

The drift term in Theorem \ref{Thm SIV-FA} is a ratio of a quadratic
polynomial in $\Delta$ divided by the square root of another quadratic
polynomial in $\Delta$. Under generic covariance structures, the leading
quadratic term in the numerator dominates and the drift grows linearly in
$|\Delta|$. In that case, the LM statistic separates the null and alternative
as $|\Delta|\to\infty$.

However, this need not occur. If the coefficient on the leading quadratic term
vanishes, then the numerator grows only linearly in $\Delta$ while the
denominator grows proportionally to $|\Delta|$. Consequently, the drift
remains bounded as $|\Delta|\to\infty$.

This motivates the following definition.

\medskip

\noindent\textbf{Assumption ID (Impossibility Design).}
\[
\mu^{\prime}\Sigma_{11}^{-1}\Sigma_{21}\Sigma_{11}^{-1}\mu= 0.
\]

\medskip

Under Assumption ID, the quadratic term in $\Delta^{2}$ in the numerator of
the LM drift vanishes. As a result, the noncentrality parameter of the LM
statistic does not diverge even when the null and alternative are arbitrarily
well separated in total variation distance.

We now characterize when such designs can occur. A data generating process is
an impossibility design if there exists $\mu\neq0$ satisfying Assumption ID.
Since $\mu$ represents the standardized first-stage coefficients, this is a
purely algebraic restriction on the reduced-form covariance structure. Define
\[
A = \Sigma_{11}^{-1}\Sigma_{21}\Sigma_{11}^{-1}.
\]
Because $A$ need not be symmetric, we consider its Hermitian part.

\begin{proposition}
\label{orthogonality and spectral decomposition} Let $A$ be a $k\times k$
matrix and define its Hermitian part:
\[
H=\frac{A+A^{\prime}}{2}.
\]
Then there exists $\mu\neq0$ such that $\mu^{\prime}A \mu=0$ if and only if the
convex hull of the spectrum of $H$ contains zero.
\end{proposition}

Proposition \ref{orthogonality and spectral decomposition} gives a geometric
characterization of impossibility designs. The condition
\[
\mu^{\prime}A\mu= 0
\]
admits a nontrivial solution if and only if the Hermitian part $H$ is not
definite. In particular, impossibility designs arise whenever $H$ has
eigenvalues of opposite sign or zero lies in the convex hull of its spectrum.

Thus impossibility designs are not knife-edge choices of $\mu$. They reflect
the covariance geometry encoded in $\Sigma$ and $\mu$. Whenever the reduced-form covariance structure is sufficiently nonorthogonal, the LM drift can be bounded.

\begin{corollary}

\label{cor:LM_ID} Under Assumptions SIV FA, NA, and ID,
\begin{align*}
&\text{LM}_{1}  - \frac{ \Delta \mu^{\prime} \Sigma_{11}^{-1} \mu} { \left(\mu^{\prime} \Sigma_{11}^{-1} \mu+
\Delta^{2} \mu^{\prime} \Sigma_{11}^{-1} \Sigma_{12} \Sigma_{11}^{-1}
\Sigma_{21} \Sigma_{11}^{-1} \mu \right) ^{1/2} } \\
\rightarrow_{d} & N \left( 0,\, 1+\|\gamma\|^{-2}\Delta^{2} \pi'
D^{1/2}\Sigma_{11}^{-1/2} M_{\gamma}\Sigma_{11}^{-1/2}
(\Sigma^{22})^{-1} \Sigma_{11}^{-1/2} M_{\gamma}\Sigma_{11}^{-1/2}
D^{1/2}\pi \right).
\end{align*}
\end{corollary}

Under Assumption ID, as $|\Delta| \to \infty$, the LM drift converges to the finite limit
\begin{equation}
\overline{m} = \frac{ \mu^{\prime}\Sigma_{11}^{-1}\mu}{ \left(  \mu^{\prime}
\Sigma_{11}^{-1} \Sigma_{12} \Sigma_{11}^{-1} \Sigma_{21} \Sigma_{11}^{-1}
\mu\right) ^{1/2} }.\label{eq:LM_bound}%
\end{equation}

This bound can be made arbitrarily small by replacing $\mu$ by $\eta\cdot\mu$ with $\eta$ small while
maintaining the impossibility design restriction. In contrast, the AR
statistic has noncentrality parameter
\[
\Delta^{2}\mu^{\prime}\Sigma_{11}^{-1}\mu,
\]
which diverges as $|\Delta|\to\infty$ whenever $\|\Sigma_{11}^{-1/2}\mu\|$ is
bounded away from zero. Hence there exist designs in which AR strongly
separates the null and alternative while LM remains weak. This separation is
structural and reflects the covariance geometry determined jointly by the
instrument design and the first-stage coefficients, as encoded in $H$.

\subsection{Implications for CQLR}

Conditional linear combination tests weight AR and LM as functions of $T$.
When LM becomes nearly uninformative, any procedure that mixes LM information
may lose power relative to procedures that rely primarily on AR. Proposition
S.1 in the online supplement provides an illustrative example in
which the LM statistic is asymptotically ancillary while the noncentrality of
AR is bounded away from zero. In this example, CLC procedures cannot outperform
a test that behaves essentially like AR. Since AR-type procedures are not
efficient under strong identification, this creates a sharp separation between
CLR and AR--LM based tests in these designs.

The same logic applies to CQLR because \citet{Andrews16} shows that CQLR is a
special case of a CLC test. Moreover, we can find a sharper bound for CQLR directly.

\begin{proposition}
\label{Prop QLR} For the QLR statistic in \eqref{QLR}: (i) if $r(T)\to\infty$,
then $QLR\to LM$; (ii) if $r(T)>\text{AR}$, then
\[
QLR \le\frac{\text{LM}\cdot r(T)}{r(T)-\text{AR}} = \frac{\text{LM}%
}{1-\text{AR}/r(T)}.
\]

\end{proposition}

When $r(T)$ diverges, the CQLR statistic collapses to LM. In such designs,
CQLR inherits the same power limitations as LM.

\begin{figure}[ptbh]
\caption{Power curves for the impossibility design with $\alpha=0.001$,
$k=10$, and $\lambda=100$.}%
\label{fig:power_comparison}%
\centering
\begin{subfigure}[b]{0.45\textwidth}
\centering
\includegraphics[width=\textwidth,trim=15mm 7mm 15mm 9mm,clip]
{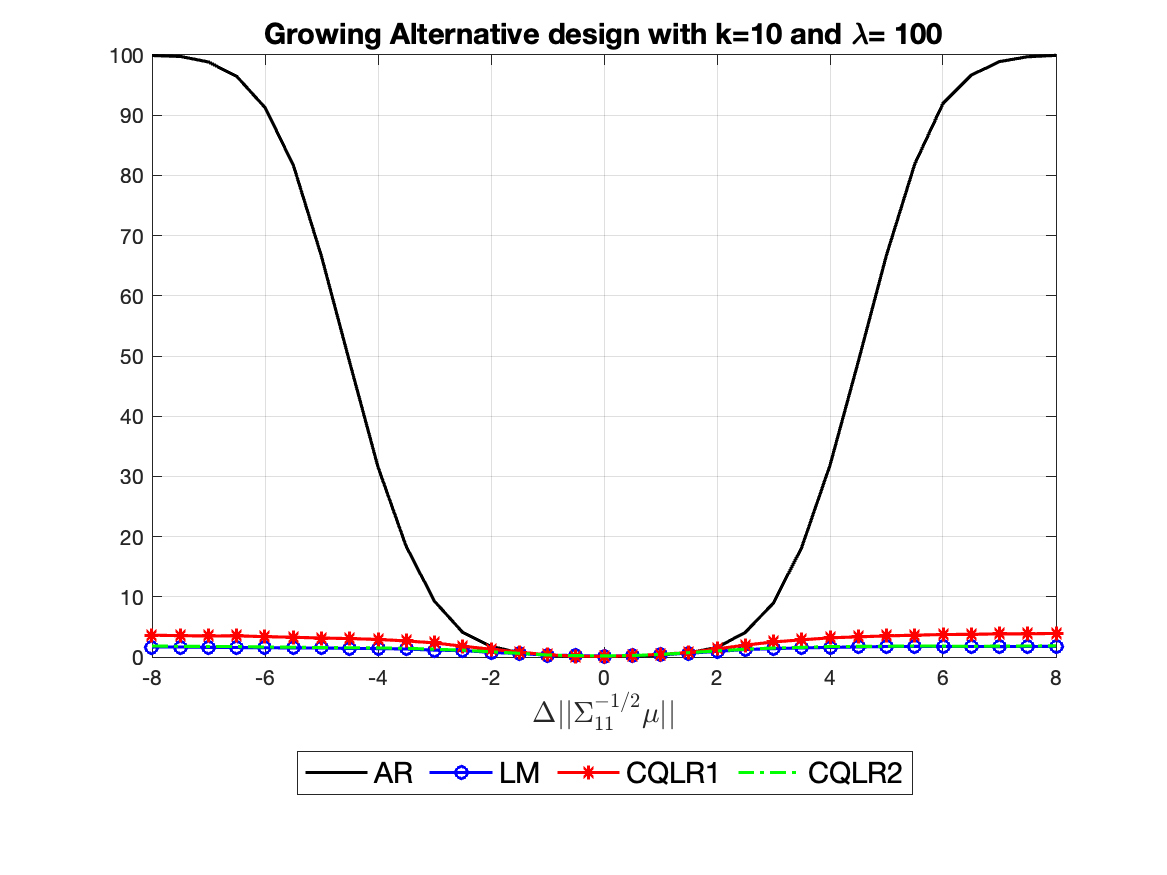}
\end{subfigure}
\hfill\begin{subfigure}[b]{0.45\textwidth}
\centering
\includegraphics[width=\textwidth,trim=15mm 7mm 15mm 9mm,clip]
{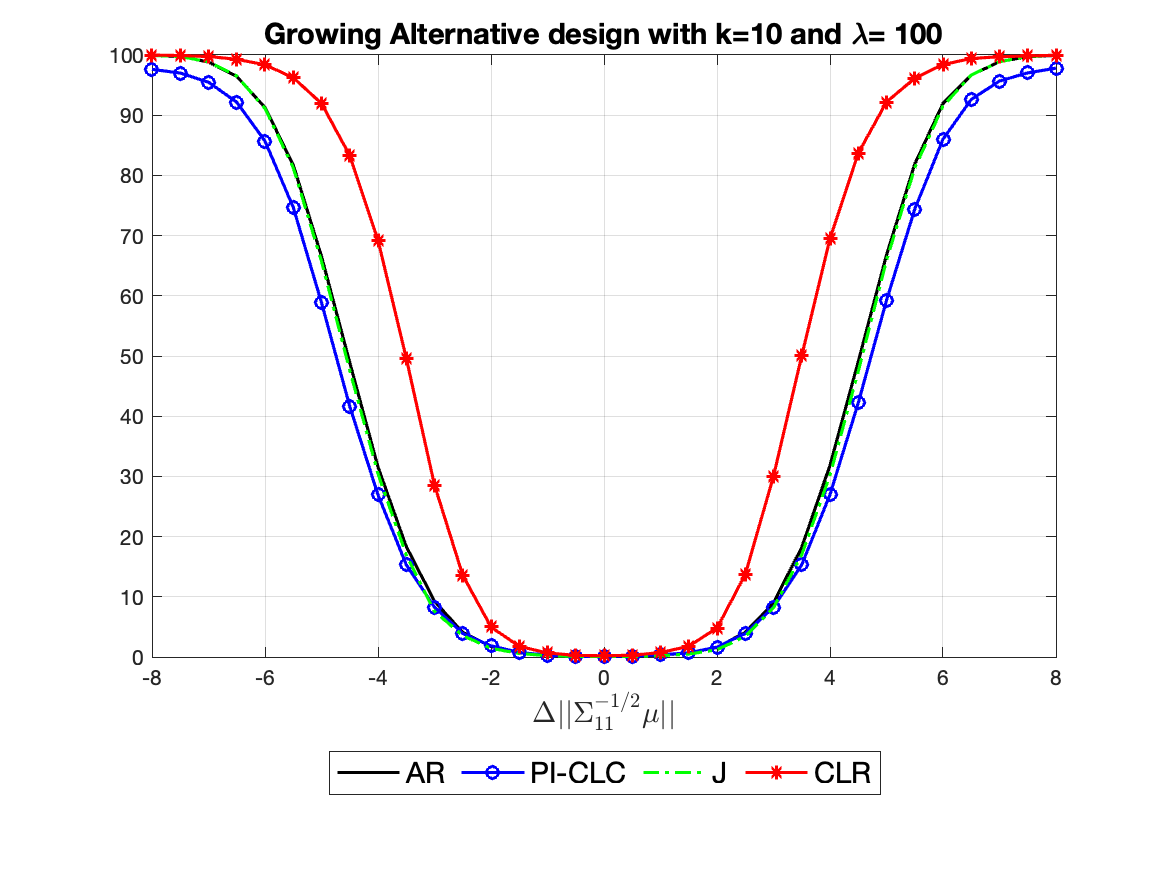}
\end{subfigure}
\end{figure}

Figure \ref{fig:power_comparison} illustrates the mechanism behind the impossibility designs and shows that LM and CQLR can fail even when the null and alternative distributions are nearly perfectly distinguishable. We assume normally distributed errors and set $\Sigma_{11}$ and $\Sigma_{22}$ proportional to the identity matrix and $\Sigma_{12}$ proportional to the anti diagonal identity matrix (that is, its $(i,j)$ element is a positive constant when $i+j=k+1$ and zero otherwise). The proportionality constants are chosen so that $\Sigma_{0}$ is positive definite. This simple covariance structure is chosen for transparency and satisfies the conditions for an impossibility design described above. This design is used only to illustrate the mechanism. The theory in the previous section shows that such failures arise for a nontrivial set of covariance structures. We take $k=10$ and set $\mu=\lambda^{1/2}e_{1}$, where $e_{1}$ denotes the first unit vector. The parameter $\lambda$ is a population $F$ statistic robust to HAC errors and measures instrument strength. We set $\lambda=100$.

We choose a very small nominal size $\alpha=0.001$ to illustrate a case in which the null and alternative are nearly perfectly distinguishable in total variation distance. The testing problem is therefore essentially trivial and the CLR power-size gap is close to one. In contrast, the LM and CQLR tests can fail to exploit this separation and may have power close to their nominal size despite the fact that the null and alternative are nearly perfectly distinguishable. The AR test separates the null and alternative but is dominated by CLR.

Section S-3 in the supplement considers several additional cases and parameter configurations for power comparisons, including impossibility design (ID) setups and near-ID setups, different values of $\alpha$, different numbers of instruments $k$, and different levels of instrument strength.

\section{Empirical Application}

\label{regions}

Section \ref{sec:LMbounds} shows that under impossibility designs the LM
noncentrality parameter can remain bounded even when the null and alternative
are well separated in total variation distance. A natural question is whether
such designs are empirically relevant or merely theoretical curiosities.

In practice, we do not observe $\mu$ directly. Instead, we estimate $\pi$ and
hence $\mu=(Z^{\prime} Z)^{1/2} \pi$. Under weak instruments, we may write
$\pi=h_{\pi}/\sqrt{n}$. Because $h_{\pi}$ is not consistently estimable, even
large samples do not allow us to determine with certainty whether the data
generating process lies on or near an impossibility design.

The empirically relevant question is therefore geometric. Does a standard
confidence region for $\mu$ intersect the set of impossibility designs? If so,
then parameter values consistent with the data can lie in regions where LM and
CQLR lose power, even though the underlying testing problem is not
intrinsically difficult.

We study the intersection between a $1-\alpha$ confidence set for $\mu$ and
the impossibility design restriction. In particular, we consider
\begin{equation}
(\widehat{\mu}-\mu)^{\prime}\Sigma_{22}^{-1}(\widehat{\mu}-\mu)\leq
q_{1-\alpha}\left(  k\right)  \quad\text{and} \quad\mu^{\prime}\Sigma
_{11}^{-1}\Sigma_{21}\Sigma_{11}^{-1}\mu=0.\label{semi-algebraic set}%
\end{equation}
The set described in \eqref{semi-algebraic set} is defined by a polynomial
inequality and a polynomial equality, and hence is semi algebraic. General
results such as the \textit{Positivstellensatz} give conditions under which a
semi algebraic set is nonempty; see \citet{Krivine64} and \citet{Stengle74}.
However, verifying those conditions is not convenient in applications. We
instead exploit the specific quadratic structure of \eqref{semi-algebraic set}
and reduce the question of feasibility to a simple optimization problem:
\begin{equation}
\min_{\mu}(\widehat{\mu}-\mu)^{\prime}\Sigma_{22}^{-1}(\widehat{\mu} -\mu)
\quad\mathrm{s.t.} \quad\mu^{\prime}H\mu=0,\label{obj}%
\end{equation}
where the Hermitian matrix is
\begin{equation}
H=\frac{\Sigma_{11}^{-1}\Sigma_{12}\Sigma_{11}^{-1} +\Sigma_{11}^{-1}%
\Sigma_{21}\Sigma_{11}^{-1}}{2}.
\end{equation}
We call the minimum value of the objective function in \eqref{obj} the
\emph{confidence bound}, because it is the smallest confidence set cutoff such
that the confidence region intersects the impossibility design. If the
confidence bound exceeds $q_{1-\alpha}(k)$, then the intersection of the
$1-\alpha$ confidence set and the impossibility design is empty.

If the convex hull of the spectrum of $H$ does not contain zero, then
trivially the only solution to \eqref{obj} is $\widetilde{\mu}=0$. Otherwise,
a solution $\widetilde{\mu}$ satisfies the first order condition
\begin{equation}
\left( \Sigma_{22}^{-1}+\kappa H\right) \widetilde{\mu} =\Sigma_{22}%
^{-1}\widehat{\mu},\label{foc}%
\end{equation}
where $\kappa$ is a Lagrange multiplier.

The matrix $\Sigma_{22}^{-1}+\kappa H$ may not be invertible. If it is
invertible, then
\begin{equation}
\widetilde{\mu} =\left( \Sigma_{22}^{-1}+\kappa H\right) ^{-1} \Sigma
_{22}^{-1}\widehat{\mu}.\label{foc invertible}%
\end{equation}
In this case, the constraint in \eqref{obj} yields the scalar constraint
equation
\begin{equation}
\widehat{\mu}^{\prime}\Sigma_{22}^{-1/2} \left( I+\kappa\overline{H}\right)
^{-1} \overline{H} \left( I+\kappa\overline{H}\right) ^{-1} \Sigma_{22}%
^{-1/2}\widehat{\mu}=0,\label{constraint}%
\end{equation}
where $\overline{H}=\Sigma_{22}^{1/2}H\Sigma_{22}^{1/2}$. In ridge regression
the left-hand side of the corresponding constraint is decreasing in $\kappa$,
so that the solution for $\kappa$ is unique.\footnote{We refer the reader to
see \citet{DraperNostrand79} for further details.} Here, because $H$ can have
eigenvalues of opposite signs, the left-hand side of \eqref{constraint} need
not be monotonic. In practice, we solve for all values of $\kappa$ numerically
and then select the $\widehat{\kappa}$ that minimizes the objective
function.\footnote{Section S-4 in the online
supplement provides more details on our search algorithm.} For any such
$\kappa$, the objective value can be written as
\begin{equation}
(\widehat{\mu}-\widetilde{\mu})^{\prime}\Sigma_{22}^{-1} (\widehat{\mu
}-\widetilde{\mu}) =\kappa^{2}\widehat{\mu}^{\prime}\Sigma_{22}^{-1/2}
\overline{H} \left( I+\kappa\overline{H}\right) ^{-2} \overline{H} \Sigma
_{22}^{-1/2}\widehat{\mu}.
\end{equation}

\begin{table}[ptbh]
\caption{Confidence bounds}%
\label{confboundtab}%
\centering
\begin{tabular}
[c]{lccc}%
\toprule & AR noncentrality ($\Delta=1$) & Confidence Bound & $F$-statistic
($H_{0}:\mu=0$)\\
& $\widetilde{\mu}^{\prime}\Sigma_{11}^{-1}\widetilde{\mu}$ & $(\widehat{\mu
}-\widetilde{\mu})^{\prime} \Sigma_{22}^{-1}(\widehat{\mu}-\widetilde{\mu})$ &
$\widehat{\mu}^{\prime}\Sigma_{22}^{-1}\widehat{\mu}$\\
\midrule Australia & -- & -- & --\\
Canada & 46.737 & 4.838 & 5.643\\
France & 109.106 & 5.127 & 6.740\\
Germany & 479.858 & 0.929 & 7.684\\
Italy & 704.791 & 0.664 & 2.067\\
Japan & 308.218 & 6.522 & 10.448\\
Netherlands & 158.154 & 1.201 & 5.674\\
Sweden & 2807.150 & 0.010 & 5.632\\
Switzerland & 160.469 & 0.031 & 0.571\\
United Kingdom & 306.603 & 1.676 & 3.850\\
United States & -- & -- & --\\
\bottomrule\bottomrule &  &  &
\end{tabular}
\end{table}

The confidence bound admits a simple geometric interpretation. It is the squared Mahalanobis distance, measured using $\Sigma_{22}^{-1}$, from the point estimate $\widehat{\mu}$ to the nearest point in the impossibility design set. If this distance is smaller than the chi--square cutoff $q_{1-\alpha}(k)$, then the data are statistically consistent with parameter values that lie on an impossibility design. If it exceeds the cutoff, the data rule out such designs at level $\alpha$.

For practitioners, this provides a direct diagnostic. The confidence bound measures how far the estimated first stage lies from the region in which LM and CQLR may lose power. It translates the algebraic condition $\mu^{\prime}\Sigma_{11}^{-1}\Sigma_{21}\Sigma_{11}^{-1}\mu = 0$ into a computable distance comparison based on standard first-stage output. In practice, this issue can also be avoided by simply using the CLR test.

As an example, we consider the estimation of the intertemporal elasticity of substitution (IES) of \Citet{Yogo04}. He considers four instruments and three models. Like \Citet{MoreiraMoreira19}, we focus on the specification in which the dependent variable is real consumption growth and the endogenous regressor is the real stock return, and we use the same estimator of the $\Sigma$ matrix as  \cite{Andrews16} does. Of the eleven countries considered by \Citet{Yogo04}, nine have eigenvalues with opposite signs, i.e., all countries except the United States and Australia. For these countries, the necessary condition for the LM noncentrality parameter to be bounded is satisfied.

Table \ref{confboundtab} shows, for all countries that satisfy the necessary condition for an impossibility design, the AR noncentrality parameter when $\Delta=1$ (second column), the minimum value of the objective function (third column, that is, the confidence bound), and the value of the objective function at $\mu=0$ (fourth column).

The second column shows that the AR noncentrality parameter when $\Delta=1$ is
at least $46.737$. Thus, separating the null from the alternative is not
intrinsically difficult. Nevertheless, the third column shows that for all
such countries the $95\%$ confidence region intersects the impossibility
design set.

In this data $k=4$, so the cutoff for the $95\%$ confidence region is $9.49$.
Since the confidence bound is below $9.49$ for each country listed, the data
do not rule out configurations in which LM and CQLR lose power.

However, $\widetilde{\mu}$ is not the only point that belongs to this
intersection. The vector $\eta\cdot\widetilde{\mu}$ for scalar $\eta$
satisfies the impossibility design restriction and lies in the confidence
region if
\begin{equation}
\left( \widehat{\mu}-\eta .\widetilde{\mu}\right) ^{\prime} \Sigma_{22}^{-1}
\left( \widehat{\mu}-\eta .\widetilde{\mu}\right)  \leq9.49.\label{conf}%
\end{equation}
As is clear from \eqref{eq:LM_bound} in Section \ref{sec:LMbounds}, choosing $\eta$ sufficiently small
can make the upper bound for the noncentrality parameter of the LM statistic
arbitrarily small. To find the smallest $\eta$ for which $\eta.\widetilde{\mu}$
lies within the confidence set, we minimize $\eta^{2}$ subject to
\eqref{conf}. The fourth column of Table \ref{confboundtab} shows that, except
for Japan, $\eta=0$ is a solution. This implies that the LM noncentrality
parameter can be arbitrarily close to zero within the confidence set. For
Japan the restriction \eqref{conf} is binding, so that
\begin{equation}
\eta= \frac{\widetilde{\mu}^{\prime}\Sigma_{22}^{-1}\widehat{\mu}}
{\widetilde{\mu}^{\prime}\Sigma_{22}^{-1}\widetilde{\mu}} - \sqrt{ \left(
\frac{\widetilde{\mu}^{\prime}\Sigma_{22}^{-1}\widehat{\mu}} {\widetilde{\mu
}^{\prime}\Sigma_{22}^{-1}\widetilde{\mu}} \right) ^{2} - \frac{\widehat{\mu
}^{\prime}\Sigma_{22}^{-1}\widehat{\mu}-9.49} {\widetilde{\mu}^{\prime}%
\Sigma_{22}^{-1}\widetilde{\mu}} }.
\end{equation}
For Japan the solution is $\eta=0.1309$.\footnote{Instead we could search over
all $\mu$ in the confidence region satisfying the impossibility design
restriction to minimize the LM bound. The semi algebraic nature of this setup
implies that the minimization problem in \eqref{obj} is a polynomial problem
for which we can find a global minimum; see \citet{Lasserre15}. We do not
pursue this approach here because the power of the LM test is already very low
in this application.}

\begin{figure}[ptbh]
\caption{Power plot for Japan considering $\mu\in\{\protect\widetilde{\mu
},0.1309\,\protect\widetilde{\mu}\}$}%
\label{fig:power_plot_japan}%
\centering
\par
\begin{subfigure}[b]{.45\textwidth}
\centering \caption{\scriptsize $\mu=\widetilde{\mu}$ (part 1)}
\includegraphics[trim=15mm 7mm 15mm 8mm,clip,width=\textwidth]
{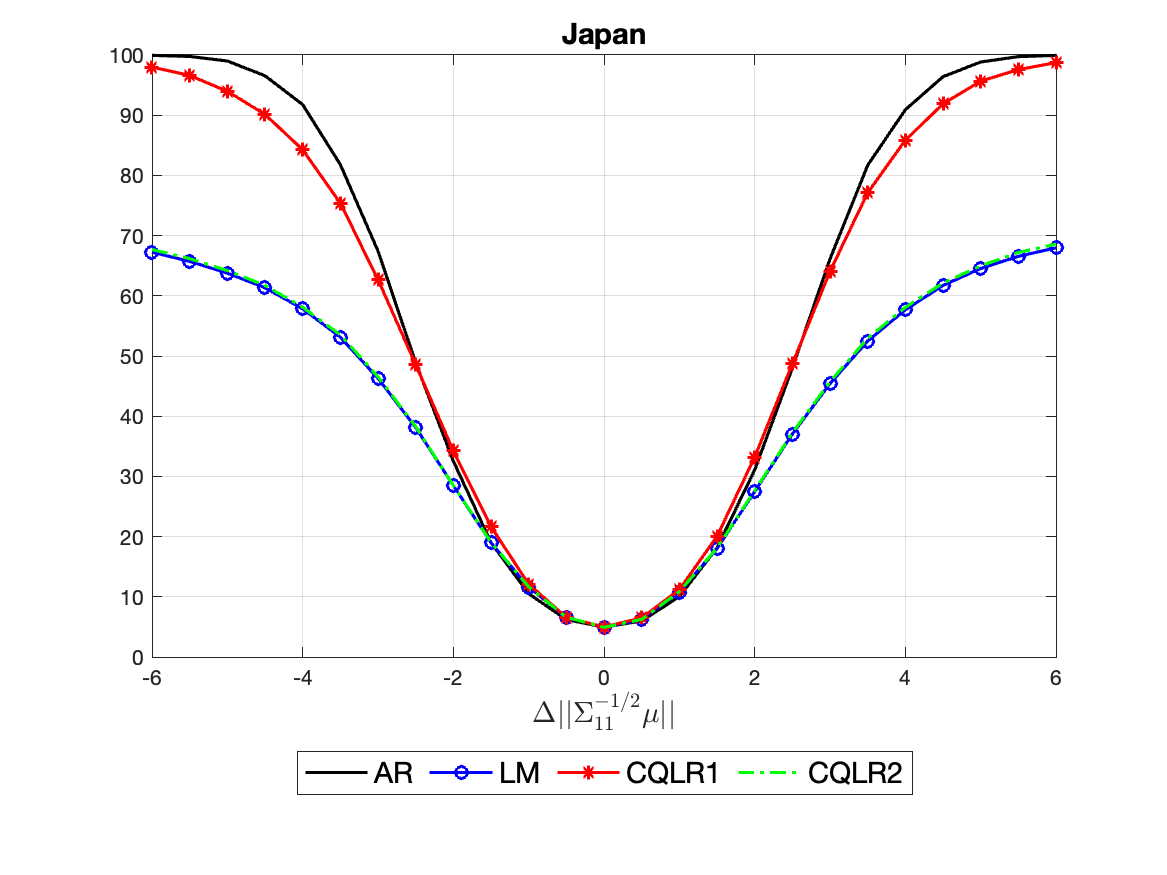}
\end{subfigure}
\begin{subfigure}[b]{.45\textwidth}
\centering \caption{\scriptsize $\mu=\widetilde{\mu}$ (part 2)}
\includegraphics[trim=15mm 7mm 15mm 8mm,clip,width=\textwidth]
{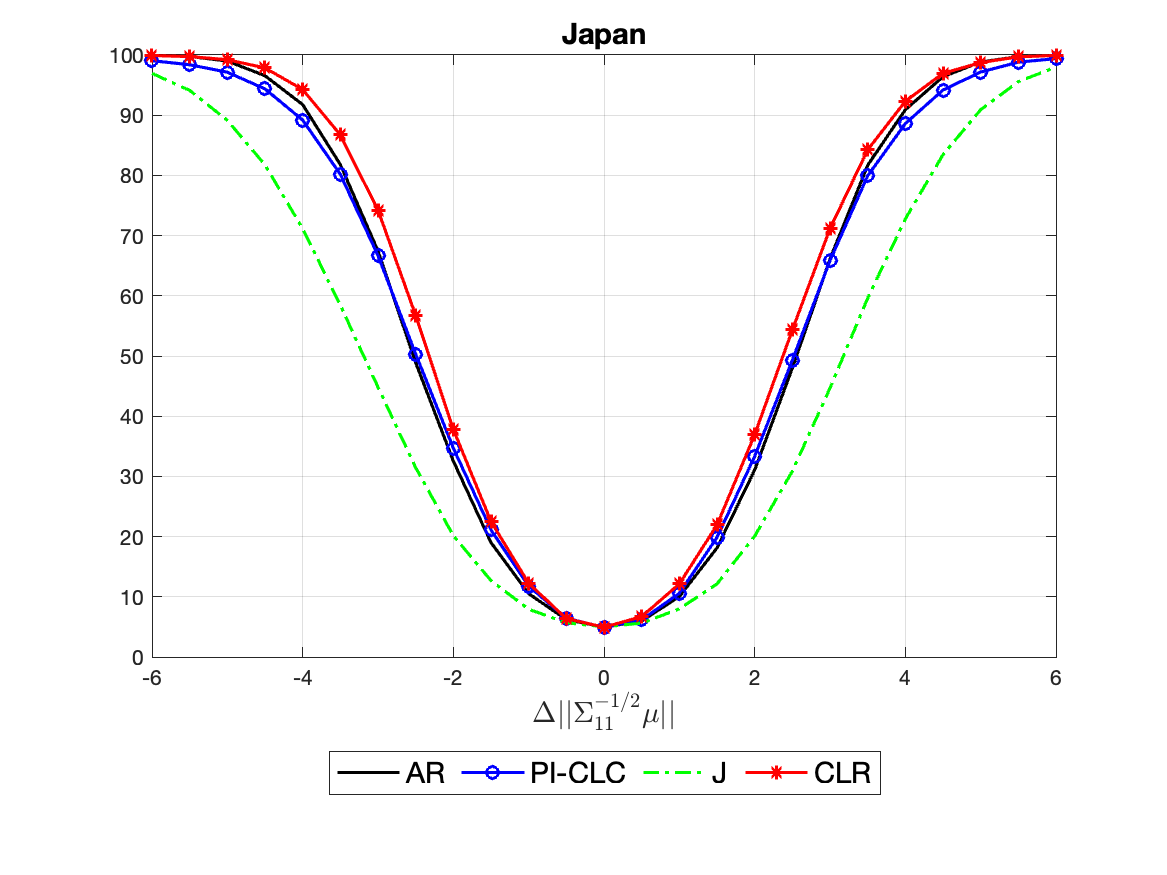}
\end{subfigure}
\par
\begin{subfigure}[b]{.45\textwidth}
\centering \caption{\scriptsize $\mu=0.1309\,\widetilde{\mu}$ (part
1)}
\includegraphics[trim=15mm 7mm 15mm 8mm,clip,width=\textwidth]
{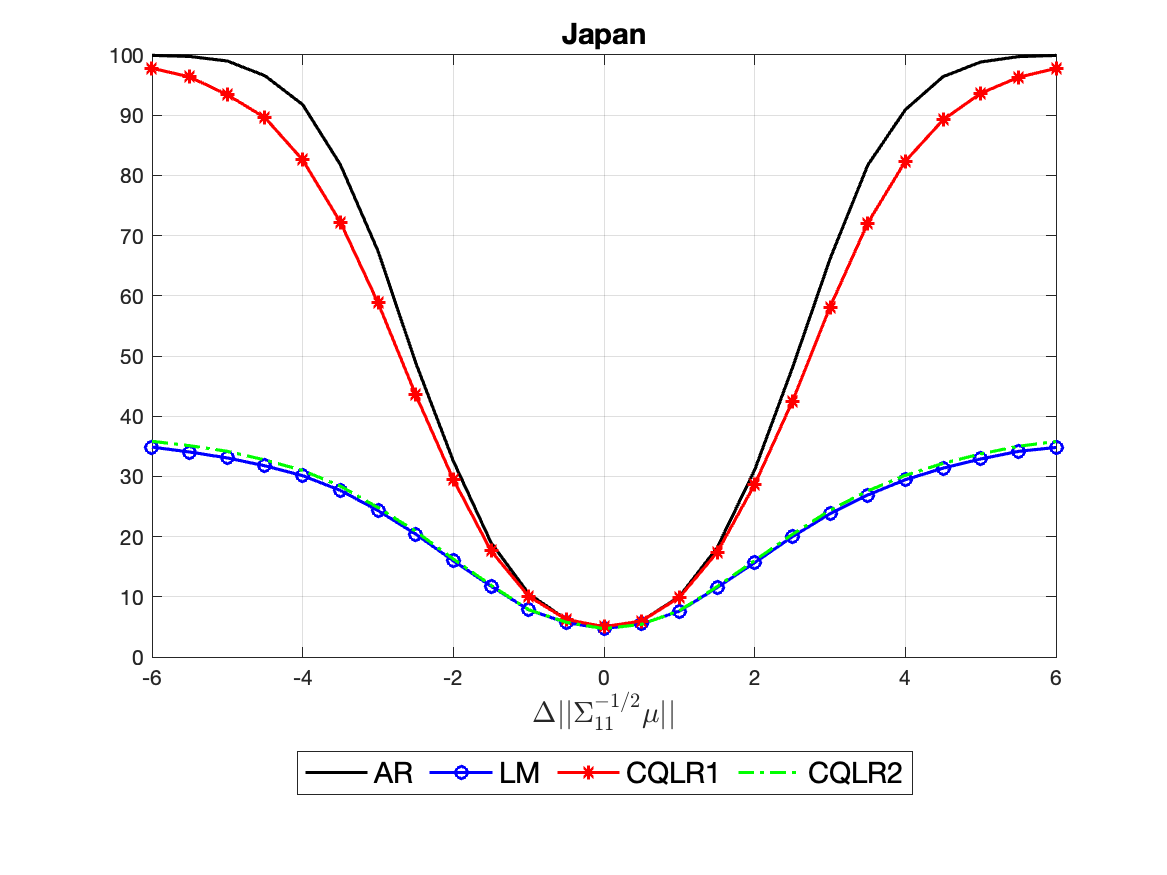}
\end{subfigure}
\begin{subfigure}[b]{.45\textwidth}
\centering \caption{\scriptsize $\mu=0.1309\,\widetilde{\mu}$ (part
2)}
\includegraphics[trim=15mm 7mm 15mm 8mm,clip,width=\textwidth]
{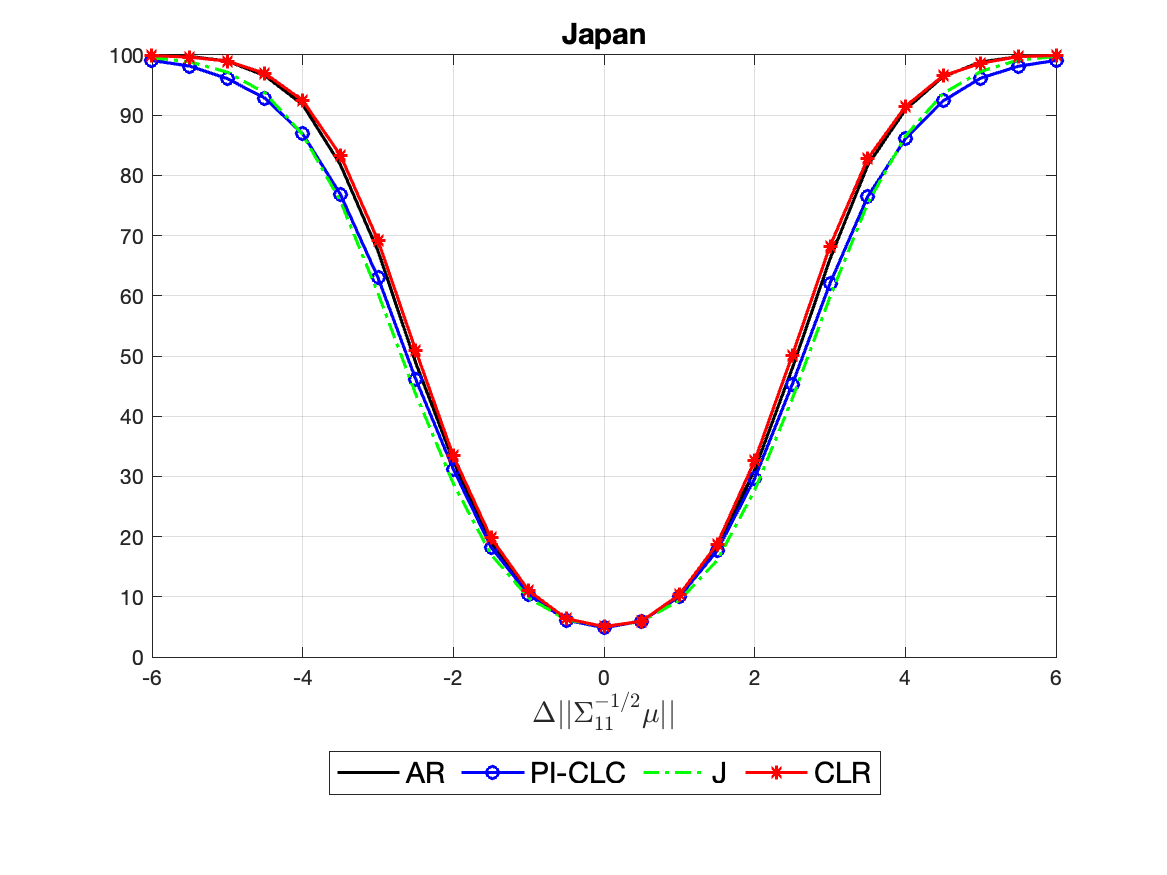}
\end{subfigure}
\end{figure}

Figure \ref{fig:power_plot_japan} shows power curves for Japan. We select this
country because it is farthest removed from the impossibility design of all countries reported in Table \ref{confboundtab}. For $\mu$ we choose the
coefficients corresponding to the confidence bound $\widetilde{\mu}$ and the
scaled value $0.1309\,\widetilde{\mu}$ on the boundary of the $95\%$
confidence set.

The power curves are consistent with the simulation results. As in the
simulations, the CLR test is not much affected by the power loss at or in the
neighborhood of an impossibility design. That conclusion holds for both
choices of $\mu$. For $k>1$ the AR test is not optimal. The LM test suffers a
substantial loss of power. Section S-5 in the
online supplement shows that $r_{2}(T)>AR$, so that Proposition \ref{Prop QLR}%
, part (ii), applies. Consequently, the CQLR2 test, for which the test
statistic behaves like the LM statistic, also suffers a substantial loss of
power. The CQLR1 dominates the CQLR2 test because the weight on the AR
component is much larger in CQLR1 than in CQLR2. The PI CLC test has power
similar to the AR test. Moreover, when the LM noncentrality shrinks toward zero, the power of the PI CLC test is bounded by the power of the $J$ overidentification test, with statistic $J = AR - LM$, as predicted by Proposition S.1 in Supplement S-1.

\begin{figure}[ptbh]
\caption{Projections of intersection of $95\%$-level set and impossibility
design (Japan)}%
\label{fig-proj}%
\begin{subfigure}{.3\textwidth}
\centering
\includegraphics[width=.9\linewidth]
{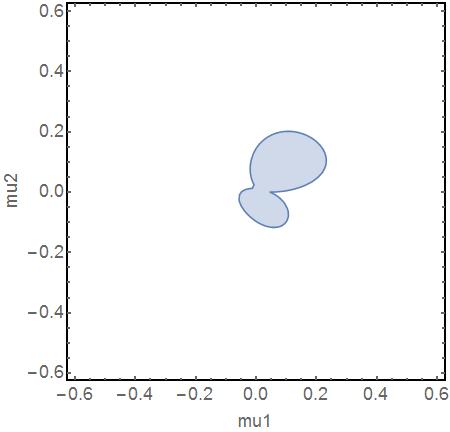}
\end{subfigure}
\begin{subfigure}{.3\textwidth}
\centering
\includegraphics[width=.9\linewidth]
{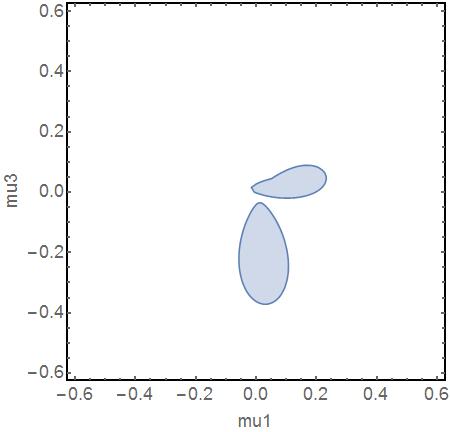}
\end{subfigure}
\begin{subfigure}{.3\textwidth}
\centering
\includegraphics[width=.9\linewidth]
{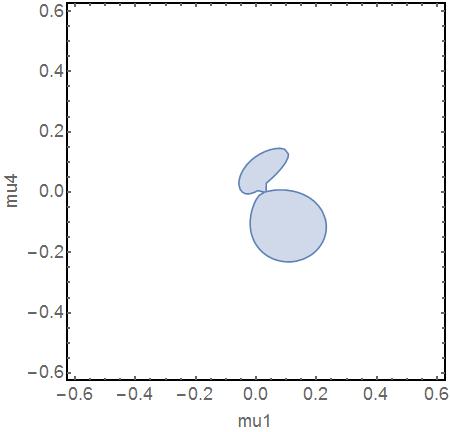}
\end{subfigure}
\newline
\par
\vspace{.3cm} \begin{subfigure}{.3\textwidth}
\centering
\includegraphics[width=.9\linewidth]
{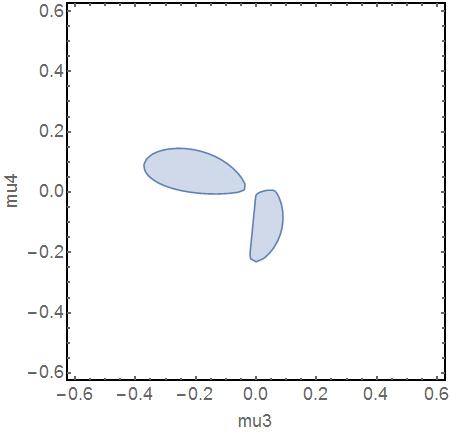}
\end{subfigure}
\begin{subfigure}{.3\textwidth}
\centering
\includegraphics[width=.9\linewidth]
{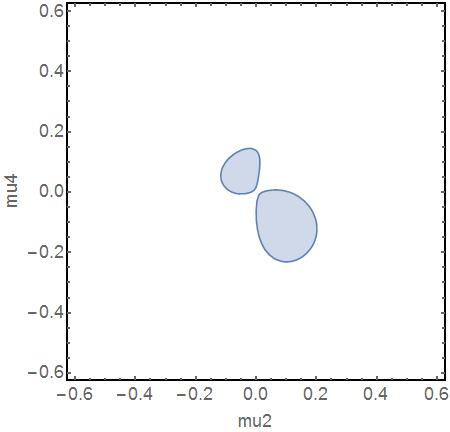}
\end{subfigure}
\begin{subfigure}{.3\textwidth}
\centering
\includegraphics[width=.9\linewidth]
{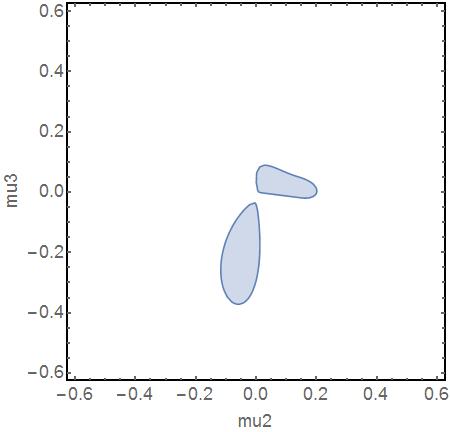}
\end{subfigure}
\end{figure}

Up to this point, we have not characterized the full set of $\mu$ that are
consistent with an impossibility design and that appear within the confidence
set. The Tarski Seidenberg theorem guarantees that projections of semi
algebraic sets are also semi algebraic sets. The Cylindrical Algebraic
Decomposition (CAD) algorithm finds these projections. Figure \ref{fig-proj}
presents all six projections in $\mathbb{R}^{2}$ of the set in $\mathbb{R}%
^{4}$ for Japan at the $95\%$ confidence level.\footnote{Throughout this paper
we take $\Sigma$ as known. If instead $\Sigma$ were estimated and allowed to
vary, the set of potentially problematic first-stage estimates would be
larger.} The graphs are close to symmetric near zero. This is not surprising
because if $\mu$ satisfies the impossibility design restriction, then so does
$-\mu$.

\section{Conclusion}

\label{sec:conclusion}

This paper characterizes the maximal attainable power--size gap in linear IV
models using total variation distance and studies which weak identification
robust procedures attain this decision theoretic frontier in the presence of
heteroskedastic or autocorrelated (HAC) errors.

Our first result establishes that the minimum total variation distance between
the convex hulls of the null and alternative models provides a sharp benchmark
for the intrinsic difficulty of the testing problem. By \citet{Kraft55}, this
distance determines the largest power--size gap achievable by any measurable
test. We show that the conditional likelihood ratio (CLR) test attains this
frontier: its power--size gap converges to one if and only if the testing
problem becomes trivial in total variation distance. Thus, whenever
statistical separation is sufficient to permit near-perfect discrimination
between the null and the alternative, the CLR test achieves it.

The contrast with AR-LM-based conditional procedures is sharp in HAC IV
models. Under fixed alternatives, retaining all terms that contribute to the
finite sample noncentrality parameter of the LM statistic reveals a class of
covariance structures, which we call impossibility designs, under which the LM
noncentrality parameter remains bounded even as the null and alternative
distributions become arbitrarily well separated in total variation distance.
In such designs, LM and CQLR can have power arbitrarily close to size despite the fact that the underlying testing problem is not intrinsically difficult. The empirical
illustration based on \citet{Yogo04} shows that confidence sets for first-stage parameters can intersect this region, so these designs arise in
empirically plausible configurations.

Taken together, these results establish a structural separation in HAC IV models. The CLR test exploits the full reduced-form information and achieves the decision theoretic frontier. Procedures constructed from AR and LM
statistics operate in a restricted information space and may fail to convert statistical separation into power. This loss is not a local artifact of asymptotic approximations. It can persist even when the null and alternative
are strongly separated.

Although our analysis is developed for HAC IV regression and extends naturally to GMM, the broader lesson concerns the evaluation of robust procedures more generally. Efficiency under standard asymptotics and robustness to weak
identification do not by themselves guarantee good performance in richer environments. When covariance structures become more complex, procedures that discard information may suffer substantial power losses while continuing to
control size. Decision theoretic benchmarks based on total variation distance provide a disciplined way to assess whether a test fully exploits the
information available in the model.

\vspace{1cm}

\textit{FGV EPGE, Praia de Botafogo, 190, 11th
floor, Rio de Janeiro, RJ 22250-040, Brasil;
mjmoreira@fgv.br}

\textit{Department of Economics, University of
Southern California, Kaprielian Hall, Los Angeles, USA, CA 90089. Electronic; ridder@usc.edu}

\textit{Department of Economics, University
of Pittsburgh, 4927 Wesley W. Posvar Hall, 230 S Bouquet St., Pittsburgh, PA
15260; sharifvaghefi@pitt.edu}

\section*{Acknowledgments}

Preliminary results of this paper were presented at seminars
organized by BU, Brown, Caltech, Harvard-MIT, PUC-Rio, University of
California (Berkeley, Davis, Irvine, Los Angeles, Santa Barbara, and Santa
Cruz campuses), UCL, USC, and Yale, at the FGV Data Science workshop, and at
conferences organized by CIREq (in honor of Jean-Marie Dufour), Harvard
University (in honor of Gary Chamberlain), Oxford University (New Approaches
to the Identification of Macroeconomic Models), and the Tinbergen Institute
(Inference Issues in Econometrics). We thank Marinho Bertanha, Leandro Gorno,
Michael Jansson, Pierre Perron, and Jack Porter for helpful comments; and
Pedro Melgar\'{e} for excellent research assistance. This study was financed
in part by the Coordena\c{c}\~{a}o de Aperfei\c{c}oamento de Pessoal de N\'
ivel Superior - Brasil (CAPES) - Finance Code 001. It was also supported in
part by the University of Pittsburgh Center for Research Computing and Data,
RRID:SCR\_022735, through the resources provided. Specifically, this work used
the H2P cluster, which is supported by NSF award number OAC-2117681.

\bibliographystyle{econometrica}
\bibliography{References}

\end{document}